\documentclass[a4paper]{jpconf}
\usepackage[utf8]{inputenc}

\usepackage{amsmath}
\usepackage{amsfonts}
\usepackage{amssymb}
\usepackage{graphicx}
\usepackage{subfig}

\newcommand{\df}{\text{d}}

\begin{document}

\title{Geometric aspects of charged black holes in Palatini theories}
\author{Gonzalo J. Olmo$^{1,2}$, D. Rubiera-Garcia$^3$ and Antonio Sanchez-Puente$^1$}
\address{$^1$ Departamento de F\'{i}sica Te\'{o}rica and IFIC, Centro Mixto Universidad de
Valencia - CSIC. Universidad de Valencia, Burjassot-46100, Valencia, Spain}
\address{$^2$ Departamento de F\'isica, Universidade Federal da
Para\'\i ba, 58051-900 Jo\~ao Pessoa, Para\'\i ba, Brazil}
\address{$^3$ Center for Field Theory and Particle Physics and Department of Physics, Fudan University, 220 Handan Road, 200433 Shanghai, China}
\ead{gonzalo.olmo@csic.es, drubiera@fudan.edu.cn, asanchez@ific.uv.es}

\begin{abstract}
  Charged black holes in gravity theories in the Palatini formalism present a number of unique properties. Their innermost structure is topologically nontrivial, representing a
 wormhole supported by a sourceless electric flux. For certain values of their effective mass and charge curvature divergences may be absent, and their event horizon may also disappear yielding a remnant. We give an overview of the mathematical derivation of these solutions and discuss their geodesic structure and other geometric properties.
\end{abstract}

\section{Introduction}

In General Relativity (GR), the curvature of the space-time is calculated using the Levi-Civita connection of the metric (Riemannian formalism). A natural way to extend GR is to allow the connection to be independent of the metric (Palatini formalism). In this way, the variation of the action will give raise to two sets of equations: one set coming from the variation with respect to the metric, and another set coming from the variation with respect to the independent connection. In exchange for the greater number of equations, these new systems turn out to be linear in the derivatives of the metric and connection, which contrasts with the second order derivatives of the metric in the usual Riemannian approach. Within the Palatini formalism, this property is preserved for Lagrangians nonlinear in curvature scalars, which allows us to consider higher curvature corrections to the gravity action without worrying about perturbative instabilities like ghosts.

We study charged black holes in this context, focusing our attention on an  action well studied in the Riemannian formalism,namely, $S=\int \left \{R+l_p^2(aR^2+R_{\mu\nu}R^{\mu\nu}) \right \} \sqrt{-g} \df^4 x$, where $l_P^2=\hbar G/c^3$ is Planck's length squared and $a$ a dimensionless constant. 
The solution for a charged black hole with this action \`{a} la Palatini was first found in \cite{or12}. This black hole solution is very similar to the Reissner-Nordstr\"om black hole of GR at scales larger than a few Planck lengths away from the center. In the Palatini solution, curvature divergences do not arise at the center but on a sphere of area $4\pi r_c^2$. However, for a certain charge-to-mass ratio curvature divergences at $r_c$ disappear. With a convenient choice of coordinates, one finds that for these configurations the metric is completely regular and the curvature invariants are finite at $r_c$. The geometry is that of a \emph{wormhole}, the radius $r_c$ being the minimum radius where the wormhole throat lies. Electric charge in this geometry arises as a topological effect resulting from the electric flux flowing through the wormhole mouth. No point-like charges are needed to generate the field. This geometry arises naturally in the Palatini context and provides a nice realization of the concept of \emph{geon} introduced by J. A. Wheeler back in 1955 \cite{Wheeler}. Given that the wormhole is a topologically non-trivial object, one can show that even configurations with curvature divergences at $r=r_c$ are indeed solutions with two sides connected at $r_c$. In this talk, we try to improve our understanding of the wormhole structure in both regular and divergent configurations by exploring the geodesic structure of these space-times and constructing Euclidean embeddings as a tool to better visualize the geometry.

\section{Solution for the metric}

We start from the action:
\begin{equation}
  S = \frac{1}{2\kappa^2}\int f(R,Q) \sqrt{-g} \df^4 x +  S_m
\end{equation}
Where $g$ is the determinant of the space-time metric $g_{\mu\nu}$, $R=g^{\mu\nu}R_{\mu\nu}$, $Q=R_{\mu\nu}R_{\alpha\beta}g^{\alpha\mu}g^{\beta\nu}$, and the matter is coupled only to the metric and not the connection. The variation of this action with respect to metric and connection (which are independent from each other) gives raise to two sets of equations (vanishing torsion)
\begin{eqnarray}
 (\partial_R f)  R_{\mu \nu} - \frac{f}{2} g_{\mu \nu} + 2 (\partial_Q f) R_{\mu \alpha} R_{\beta\nu}g^{\alpha \beta}&=& \kappa^2 T_{\mu \nu} \label{eq:metric}\\
 \nabla_\beta \left [ \sqrt{-g}((\partial_R f) g^{\mu \nu} + 2 (\partial_Q f) R^{\mu \nu} ) \right ] &=& 0 \label{eq:connection}
\end{eqnarray}

The details to solve the equations given $f(R,Q)$ and $S_m$ are given in \cite{or12}. A quick summary goes as follows: Eq. (\ref{eq:metric}) can be manipulated raising one index with the metric to show that ${P_\mu}^{\nu}\equiv R_{\mu\alpha}g^{\alpha\nu}$ can be expressed as an algebraic function of ${T_\mu}^\nu$. Being that relation a matrix equation, there might be multiple solutions. However, agreement with GR in the low curvature regime should select just one of them. Tracing over the indices of ${P_\mu}^\nu$, one finds $R={P_\mu}^\mu$, whereas $Q={P_\mu}^\alpha {P_\alpha}^\mu$. With these expressions, it is possible to obtain the functional form of $R$ and $Q$ in terms of the matter fields and, in the case of an electric field with spherical symmetry, in terms of $r$. Now, with ${P_\mu}^\nu$ written in terms of the matter fields, we can write (\ref{eq:connection}) in the form $ \nabla_\beta \left [ \sqrt{-g}g^{\mu\alpha}((\partial_R f) {\delta_\alpha}^\nu + 2 (\partial_Q f) {P_\alpha}^\nu  \right ] = 0$. Here the connection appears linearly and can be solved using elementary algebraic manipulations. A simple way to see this, consists on defining an auxiliary metric $h_{\mu\nu}$ such that the connection equation becomes $\nabla_\beta \left [\sqrt{-h} h^{\mu\nu} \right ]= 0$. This implies that $\Gamma^\alpha_{\mu\nu}$ is the Levi-Civita connection of $h_{\mu\nu}$, with $h_{\mu\nu}$ and $g_{\mu\nu}$ related by
\begin{equation} \label{eq:h-g}
h^{\mu\nu}=\frac{g^{\mu\alpha}{\Sigma_{\alpha}}^\nu}{\sqrt{\det \hat{\Sigma}}} \ , \quad
h_{\mu\nu}=\left(\sqrt{\det \hat{\Sigma}}\right){{\Sigma^{-1}}_{\mu}}^{\alpha}g_{\alpha\nu} \ ,
\end{equation}
where ${\Sigma_\alpha}^{\nu}=\left(\partial_R f \delta_{\alpha}^{\nu} +2\partial_Q f {P_\alpha}^{\nu}\right) $. With this notation, it is possible to write eq. (\ref{eq:metric}) only in terms of $h_{\mu\nu}$ (and its derivatives), $R$ and $Q$ as functions of $r$, and $T^\mu{}_\nu$. The resulting equation can be integrated to obtain $h_{\mu \nu}$,  which allows to compute $g_{\mu\nu}$ using the relations (\ref{eq:h-g}).

Before showing the metric solution for the particular case of $f(R,Q)=R+l_p^2(aR^2+Q)$, and the $T^\mu{}_\nu$ of electrovacuum fields, let us define some constants and factors that will appear: integration of the the equations will give a constant $r_S$ that is identified with the Schwarzschild radius of the black hole. The number of charges of the black hole together with the Planck length gives two characteristic lengths $r_q=N_q l_P$ (where $N_q$ is the number of charges) and $r_c=\sqrt{r_q l_P}$. With these we can define a dimensionless parameter $\delta_1$ that will characterize the behaviour of the black hole near the wormhole throat, and two factors $\sigma_\pm(r)$ that represent the non-zero components of the matrix ${\Sigma_\alpha}^{\nu}$ introduced above
\begin{equation}
 \delta_1 = \frac{1}{2}\frac{r_q^2}{r_c r_s} \qquad \sigma_\pm(r)=1\pm\frac{r_c^4}{r^4} \ .
\end{equation}
With this notation, the metric takes the form
\begin{equation}
  ds^2 = -A(r) dt^2 + \frac{1}{A(r)\sigma_+^2} dx^2 + r^2(x) d \Omega^2 \ ,
\end{equation}
where the area function $r^2(x)$ and the radial coordinate $x$ are related as follows
\begin{equation}
 x^2=r^2\sigma_-  \ , \ r^2=\frac{x^2+\sqrt{x^4+4 r_c^4}}{2} \ .
\end{equation}
The behaviour of $A(r)$ in the limit of large and small  $r$, respectively, is \cite{or12,lor,lmor1,lmor2}
\begin{equation}
\begin{array}{lll}
  A(r)\approx 1-\frac{r_S}{ r  }+\frac{r_q^2}{2r^2} + O \left ( \frac{r_c^4}{r^4} \right ) &\text{for}& r \gg r_c\\
  A(r)\approx \frac{N_q}{4N_c}\frac{\left(\delta _1-\delta _c\right) }{\delta _1 \delta _c }\sqrt{\frac{r_c}{ r-r_c} }+\frac{N_c-N_q}{2 N_c}+O\left(\sqrt{r-r_c}\right)&\text{for}& r \sim r_c
\end{array}
\end{equation}
Here $\delta_c \simeq 0.572069$ and $N_c \simeq 16.55$. The rapid decay of the corrections for $r\gg r_c$ puts forward that at distances slightly larger than $r_c$ the metric is essentially the same as in the Reissner-Nordstr\"om black hole of GR. For stellar size objects, no difference with respect to GR could be seen outside the horizon or even a few Planck lengths away from the center. However, as $r\to r_c$, the behavior changes radically and becomes strongly dependent on $\delta_1$. If $\delta_1$ is greater than $\delta_c$, there is a time-like curvature divergence. If $\delta_1$ is smaller than $\delta_c$, the divergence is space-like. For $\delta_1=\delta_c$ the geometry becomes completely smooth. Moreover, if the number of charges of the black hole is less than a critical number, $N_q<N_c$ there will be no horizon hiding the hypersurface $r=r_c$.

\begin{figure}[h]
\centering
\includegraphics[width=.4\linewidth]{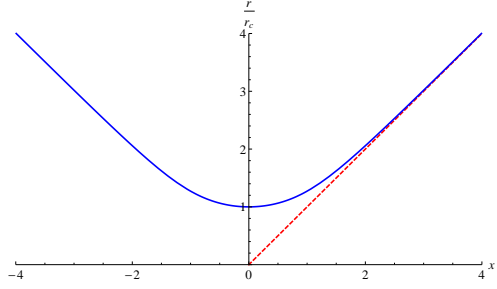}
\caption{$r$ as a function of $x$, in units of $r_c$}
\label{fig:xrgraph}
\end{figure}

Fig. \ref{fig:xrgraph} illustrates why the radial function $r(x)$ cannot be used as a valid coordinate. When $x=0$, the function $r(x)$ has a minimum and ceases to be monotonic, which prevents it from mapping the whole domain of the solution. It is important to note that the relation between $x$ and $r(x)$ is independent of the charge-to-mass ratio. In the regular case, $\delta_1=\delta_c$, nothing prevents extending the domain of $x$ to the negative axis, i.e.,  $x\in ]-\infty,+\infty[$. Given that the wormhole is a topological structure, thus insensitive to metric perturbations, the domain of $x$ must always be the same. This is further supported by the properties of the electric flux across the wormhole throat, which is independent of the charge-to-mass ratio. Moreover, the ratio of the electric flux by the area of the wormhole turns out to be a universal constant, independent of the charge and mass of the solutions. This further reinforces the topological character of the solution and the existence of {\it two sides} regardless of the existence of curvature (metric) divergences at the throat.

\section{Euclidean embeddings}

\begin{figure}[h]
\centering
\includegraphics[width=.4\linewidth]{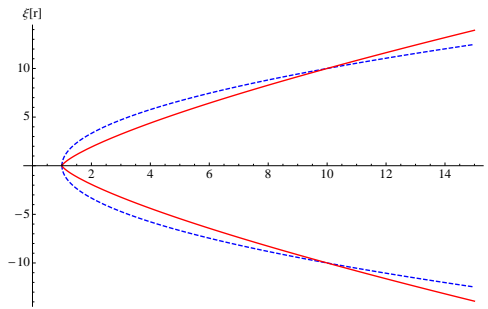}
\caption{$\xi(r)$ for the regular case (blue dotted line) and the singular case (red line)}
\label{fig:Xi}
\end{figure}

\begin{figure}[h]
\centering
\subfloat{\includegraphics[width=.4\textwidth]{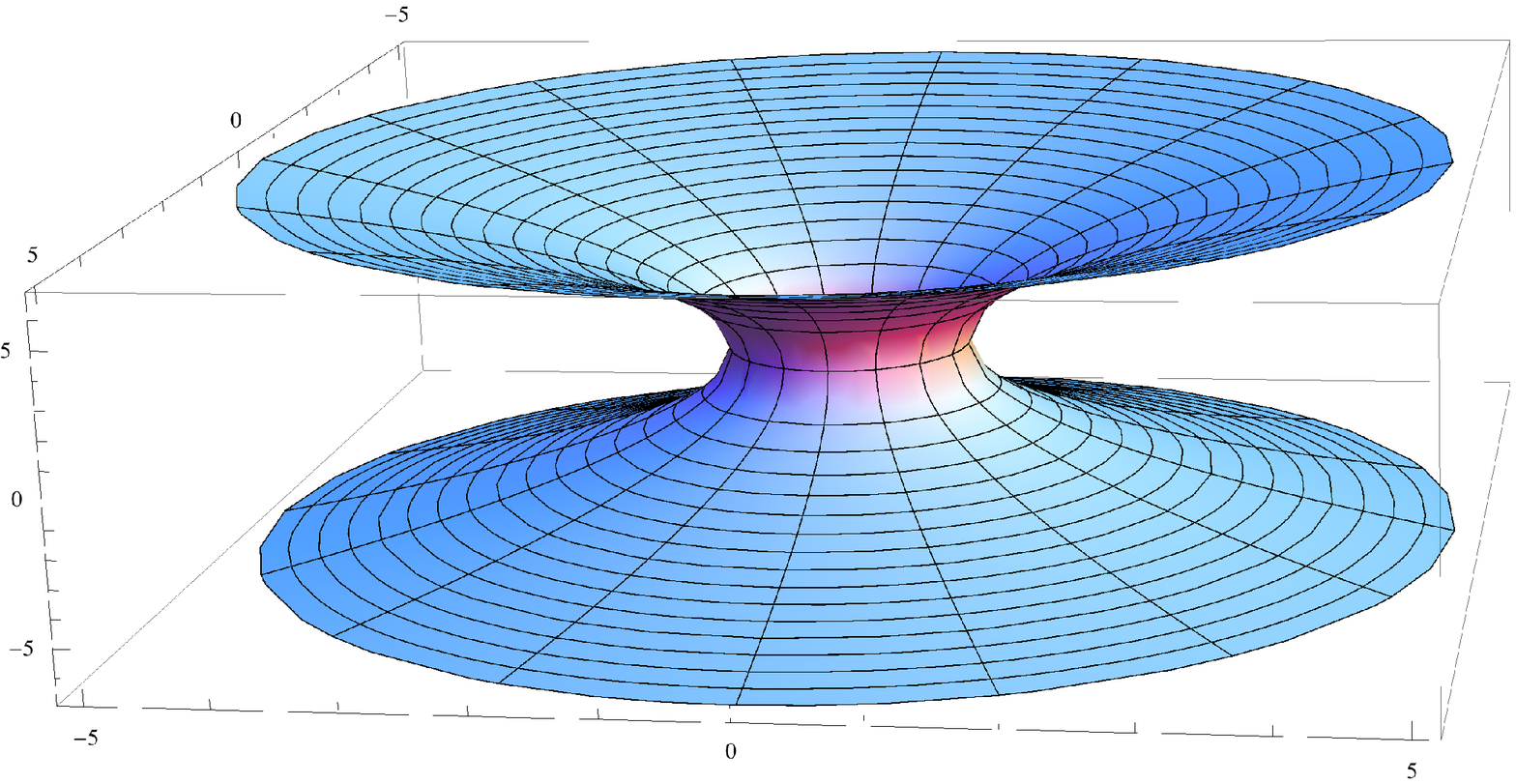}}
\subfloat{\includegraphics[width=.4\textwidth]{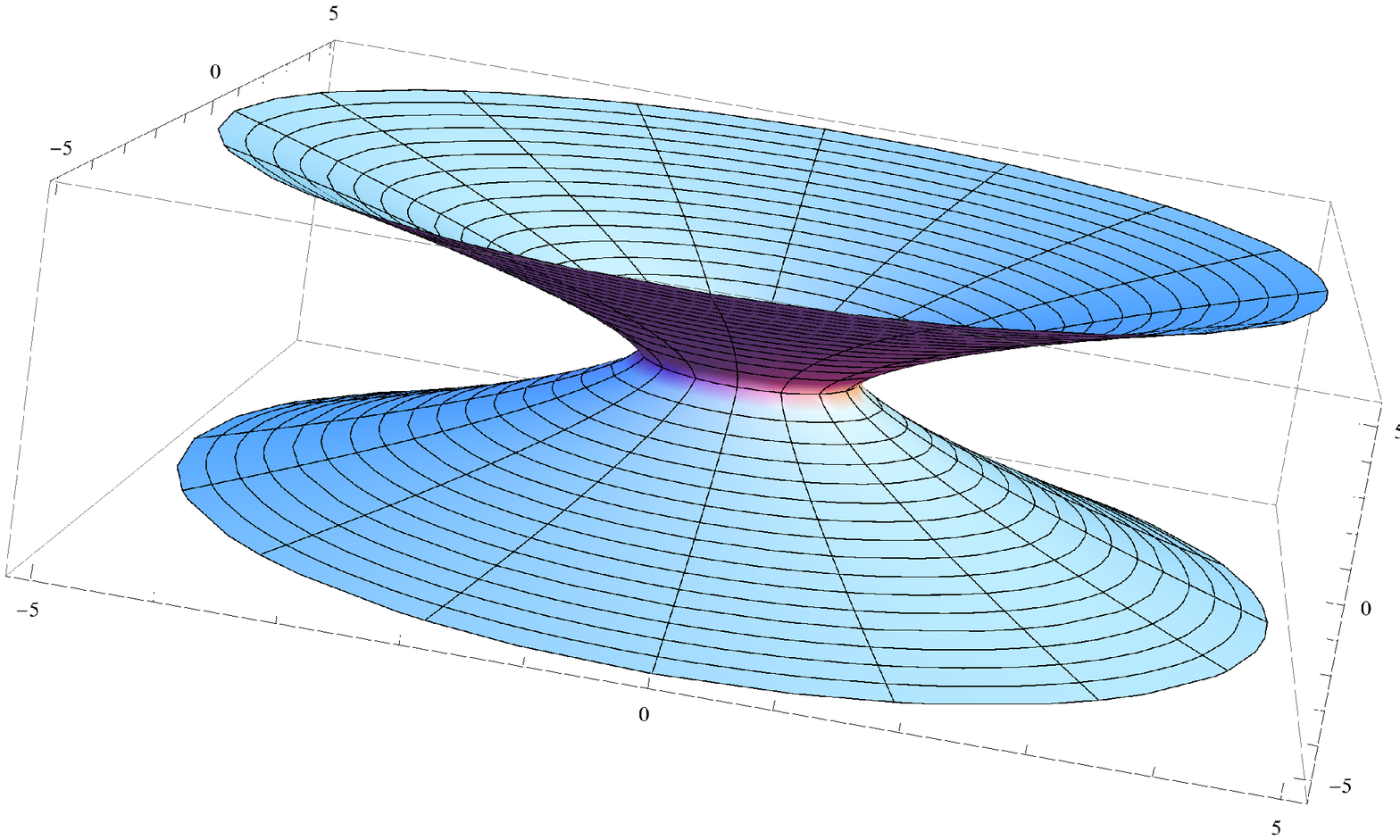}}
\caption{Embedding of a bidimensional slice of the spacetime in 3D euclidean space fo the regular and the singular solution}
\label{fig:Embeddings}
\end{figure}

The geometry changes from being smooth to having divergences if $\delta_1$ moves infinitesimally away from the value $\delta_c$. It seems shocking that such a small change can lead from something regular to something as pathological as curvature divergences are thought to be. In order to better understand the differences between such configurations, it is useful to look at the Euclidean embedding of a bidimensional slice of the spacetime ($t=constant, \theta= \pi/2$). The induced metric in this slice is
\begin{equation}\label{eq:2D}
\df l^2=\frac{1}{A\sigma_-}\df r^2+r^2\df \varphi^2 \ ,
\end{equation}
while the metric of the three-dimensional Euclidean space with cylindrical symmetry in which we want to embed this slice has the form \cite{MTW}
\begin{equation}
\df l^2=\df \xi^2+\df r^2+r^2\df \varphi^2.
\end{equation}
To find the embedding one just needs to restrict $\xi$ to be a function $\xi(r)$ so that the line element in eq. (\ref{eq:2D}) is recovered. Fig. \ref{fig:Xi} shows $\xi(r)$ for two particular cases with $\delta_1=\delta_c$ and $\delta_1>\delta_c$ (the latter representing the analogous of a naked singularity in GR). In the regular case the wormhole is smooth, whereas in the $\delta_1>\delta_c$ case, there is a kind of vertex at the wormhole throat. It is worth noting that in the $\delta_1>\delta_c$ case $r(\xi) \simeq \xi^\frac{4}{3}$ is both continuous \emph{and} differentiable everywhere. It is the second derivative of $r(\xi)$ that tends to infinity as the wormhole throat is approached, causing the divergence of curvature scalars. The embeddings are shown in Fig. \ref{fig:Embeddings}.

\section{Geodesics}

Geodesics are curves whose tangent vector is parallel transported along itself. This definition generalizes the concept of ``straight lines'' of Euclidean geometry to curved geometry. According to the Einstein Equivalence Principle, freely falling objects follow geodesics. A geodesic $x^\mu(\lambda)$ satisfies the following equation:

\begin{equation}
  \frac{\df^2  x^\mu}{\df  \lambda^2}+\Gamma^\mu_{\alpha \beta} \frac{\df x^\alpha}{\df \lambda}\frac{\df x^\beta}{\df \lambda} = 0 \ , \label{eqgeo}
 \end{equation}

In the Riemannian formalism, the affine connection  that appears in the geodesic equation is the Levi-Civita connection of the metric. However, in the Palatini formalism, the independent connection could also be used to define geodesic paths. If the theory is constructed in such a way that the connection does not couple to the matter fields (as in our case), the geodesics of freely falling objects (or of light rays) are the ones derived from the Levi-Civita connection of the metric.

Since the geometry we are dealing with is very symmetric, it is not necessary to solve the geodesic equation. Time translation symmetry gives us a conserved quantity along the geodesic $E=A\frac{\df t}{\df \lambda}$, and spherical symmetry gives us another conserved quantity, $L=r^2\frac{\df \varphi}{\df \lambda}$. The radial component can be obtaining normalizing the tangent vector to $\kappa=0$ or $\kappa = 1$ depending if it is a null or a time-like geodesic:
  \begin{equation}
\frac{1}{\sigma_+(x)}\frac{d x}{d \lambda}=\sqrt{{E^2-\underbrace{A(x)\left(\kappa+\frac{L^2}{r^2(x)}\right)}_{V(x)}}}
 \end{equation}
The motion in the radial direction is analog to the motion of a particle in a one dimensional potential $V(x)$ with energy $E^2$. In this case, the energy is squared and cannot take negative values. Therefore, wherever the potential $V(x)$ is negative, $\frac{d x}{d \lambda}$ is necessarily different from $0$ and the particle is forced to cross that region. The regions where $V(x)$ is negative correspond to the regions where $A(x)$ is negative (inside horizons) and the $x$ coordinate becomes time-like (the $t$ coordinate becomes space-like). The zeros of the potential correspond to the regions where $A(x)$ vanishes, which occurs at the black hole horizons.

Radial null geodesics are characterized by $\kappa=0$, $L=0$. The potential is null in this case and insensitive to the details of $A(x)$. In this case, light rays will follow their path crossing the wormhole unimpeded.

\begin{figure}[h]
\centering
\subfloat[$\delta_1=\delta_c$]{\includegraphics[width=.33\textwidth]{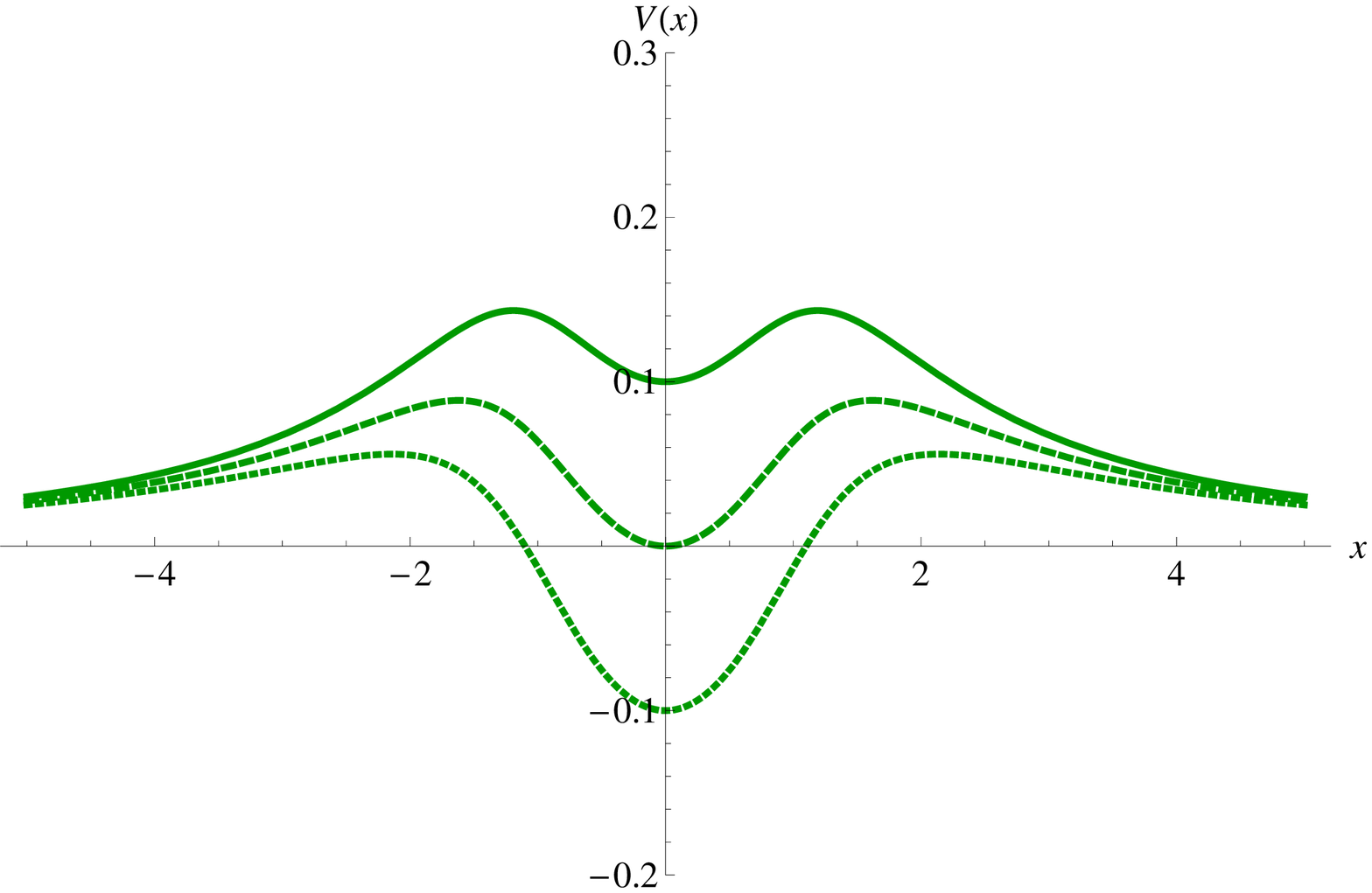}}
\subfloat[$\delta_1>\delta_c$]{\includegraphics[width=.33\textwidth]{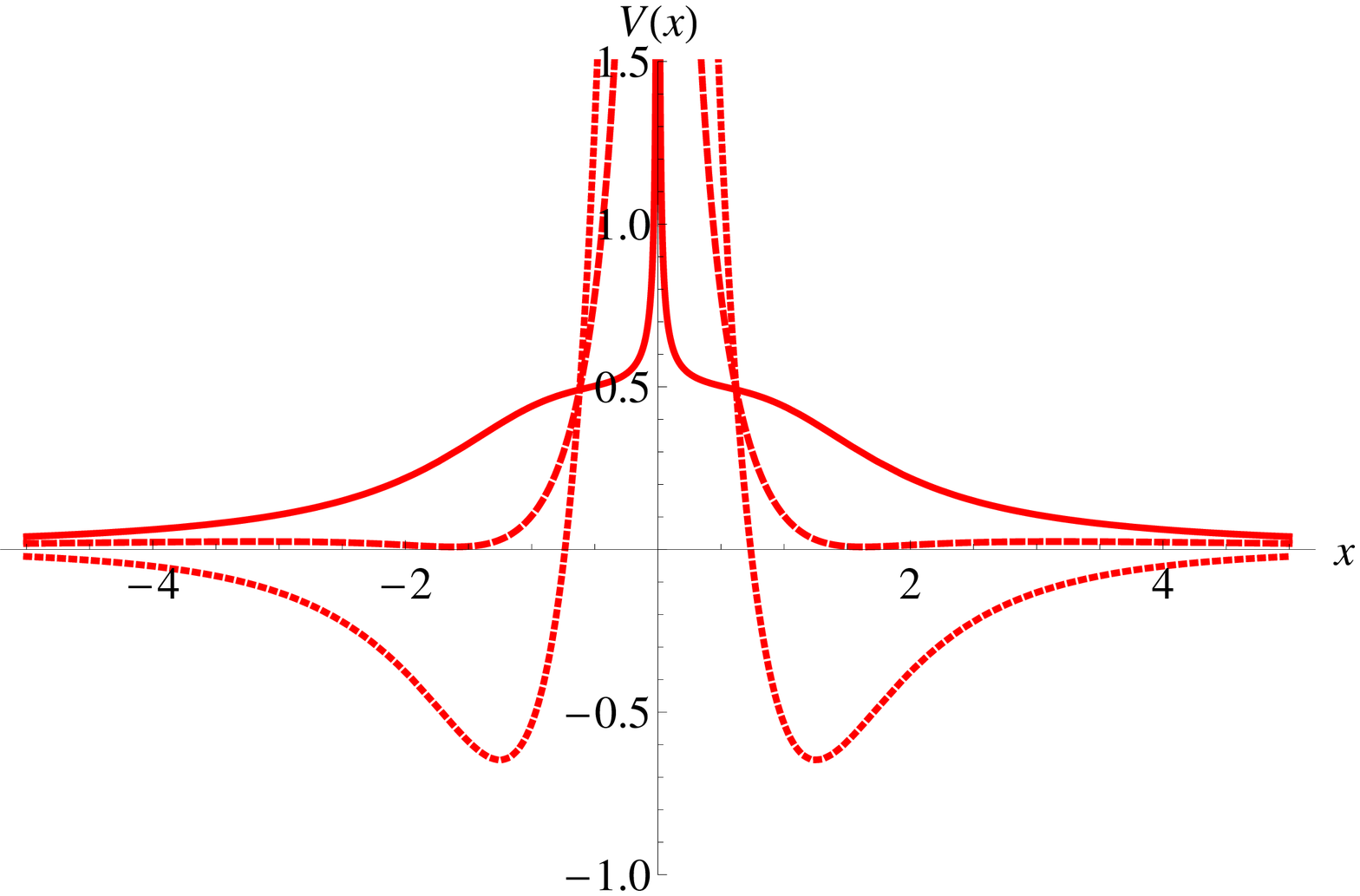}}
\subfloat[$\delta_1<\delta_c$]{\includegraphics[width=.33\textwidth]{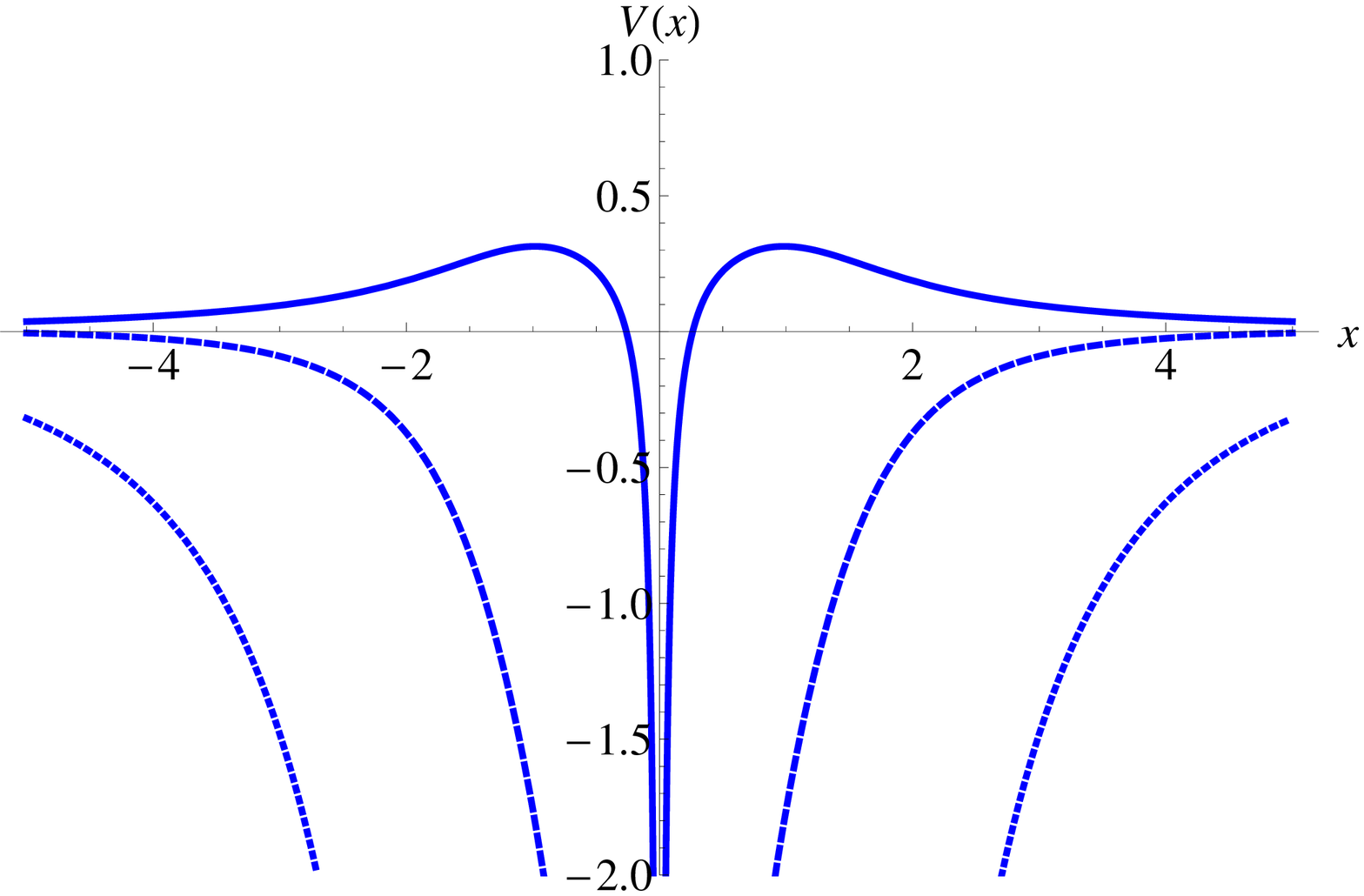}}
\caption{Potential for null geodesics with angular momentum. In each plot, the potential is shown for 3 different values for the number of charges $N_c$ (but the same $\delta_1$)}
\label{fig:NullAngular}
\end{figure}

The potential for null geodesics with angular momentum is shown in Fig. \ref{fig:NullAngular}. For large values of $r$, the geodesics feel a repulsive potential that goes like $L/r^2$. Near the wormhole, the behaviour of the potential depends on the value of $\delta_1$. If $\delta_1<\delta_c$, the potential will reach a maximum and then go to $-\infty$ as the geodesic approaches the wormhole. All geodesics with energy greater than the maximum will cross to the other side of the wormhole. If $\delta_1=\delta_c$ the potential will have a finite value at the wormhole throat and reach a maximum before that. The potential at the wormhole throat is a minimum, and there will be stable light orbits if the value of the potential is positive (which happens when $N_q<N_c$). If $\delta_1>\delta_c$, the potential will go to $+\infty$ as the
geodesic approaches the wormhole throat. As a consequence,  all geodesics will be repelled and none will cross the wormhole.

\begin{figure}[h]
\centering
\subfloat[$\delta_1=\delta_c$]{\includegraphics[width=.33\textwidth]{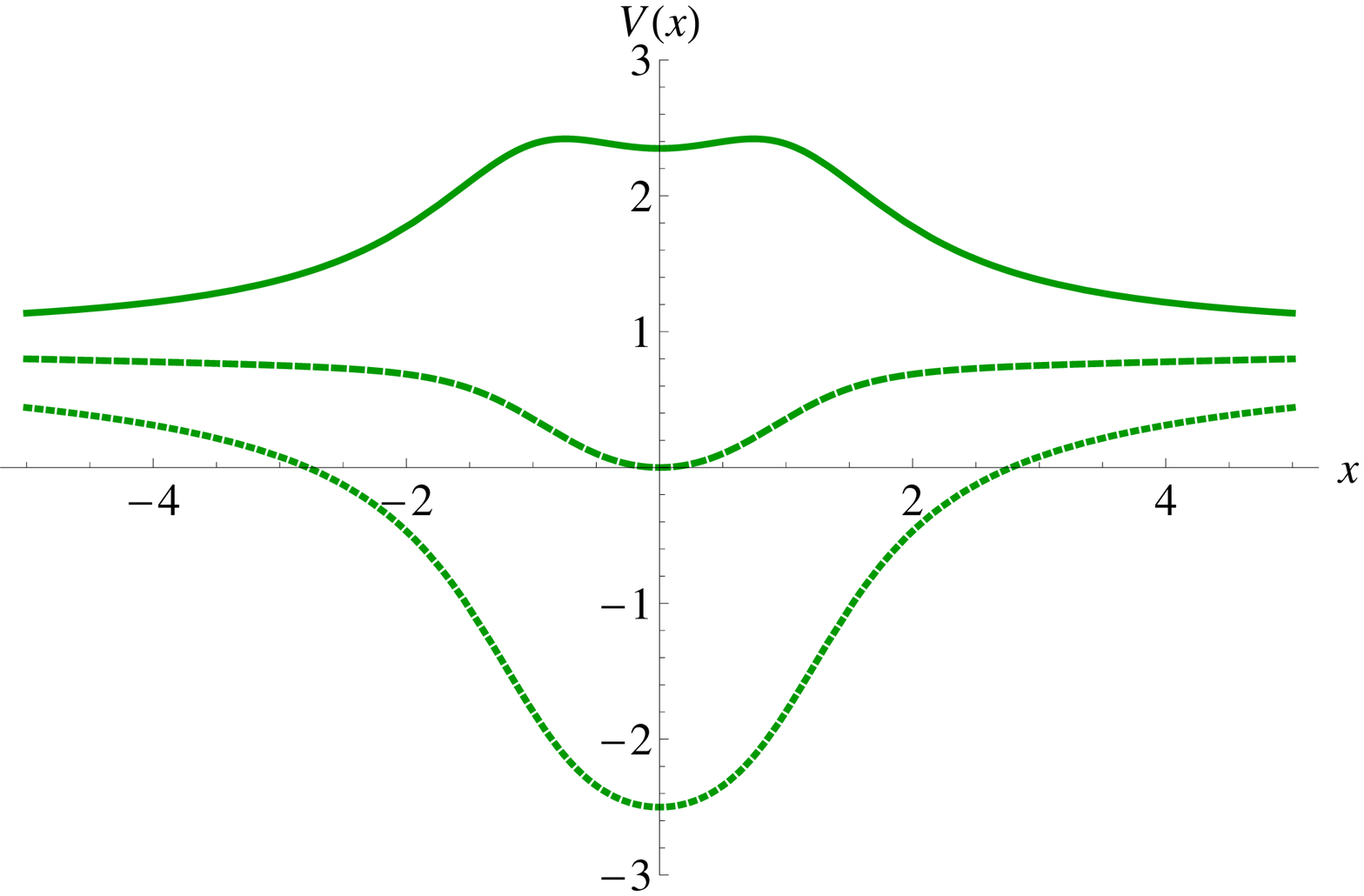}}
\subfloat[$\delta_1>\delta_c$]{\includegraphics[width=.33\textwidth]{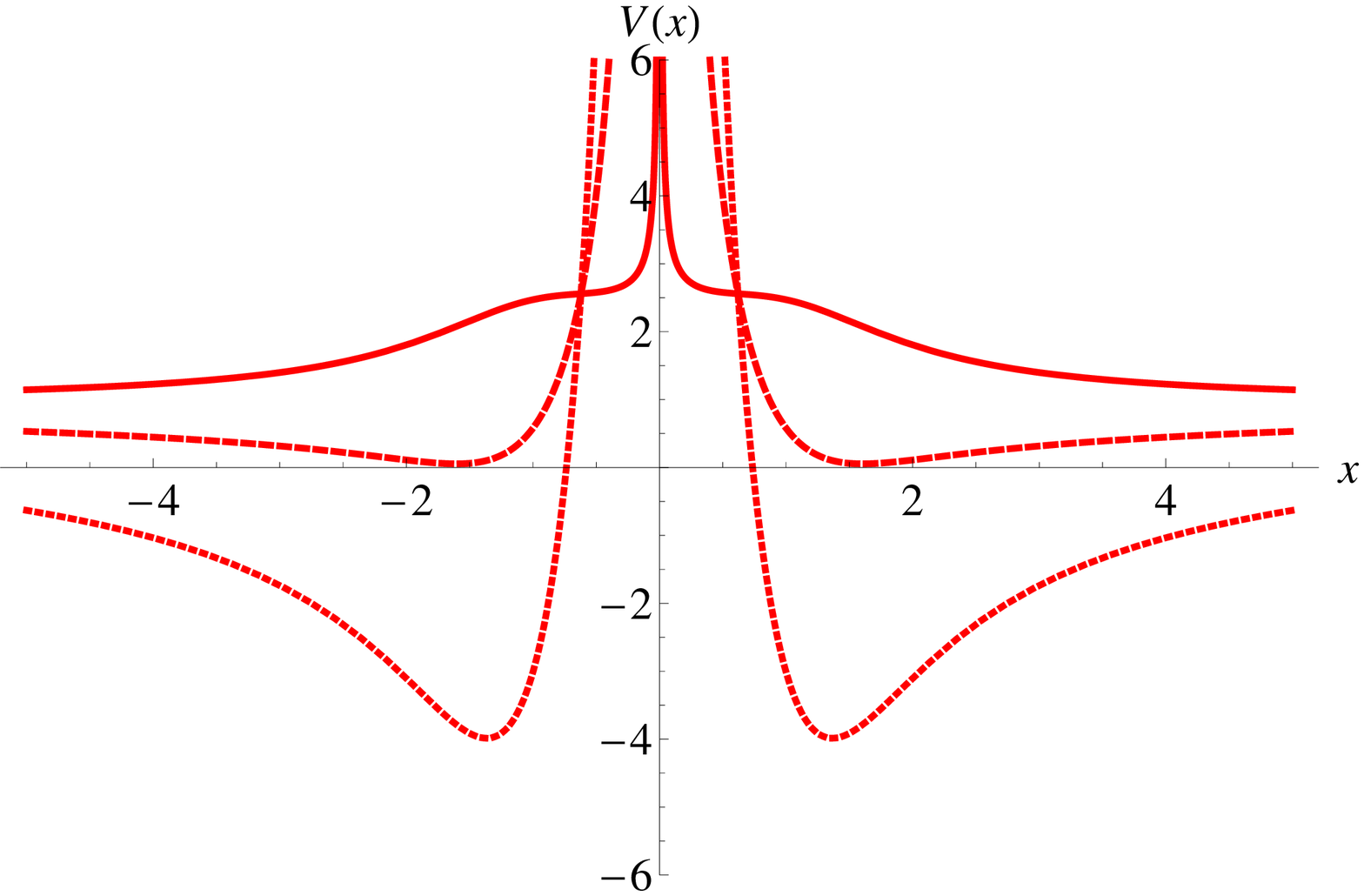}}
\subfloat[$\delta_1<\delta_c$]{\includegraphics[width=.33\textwidth]{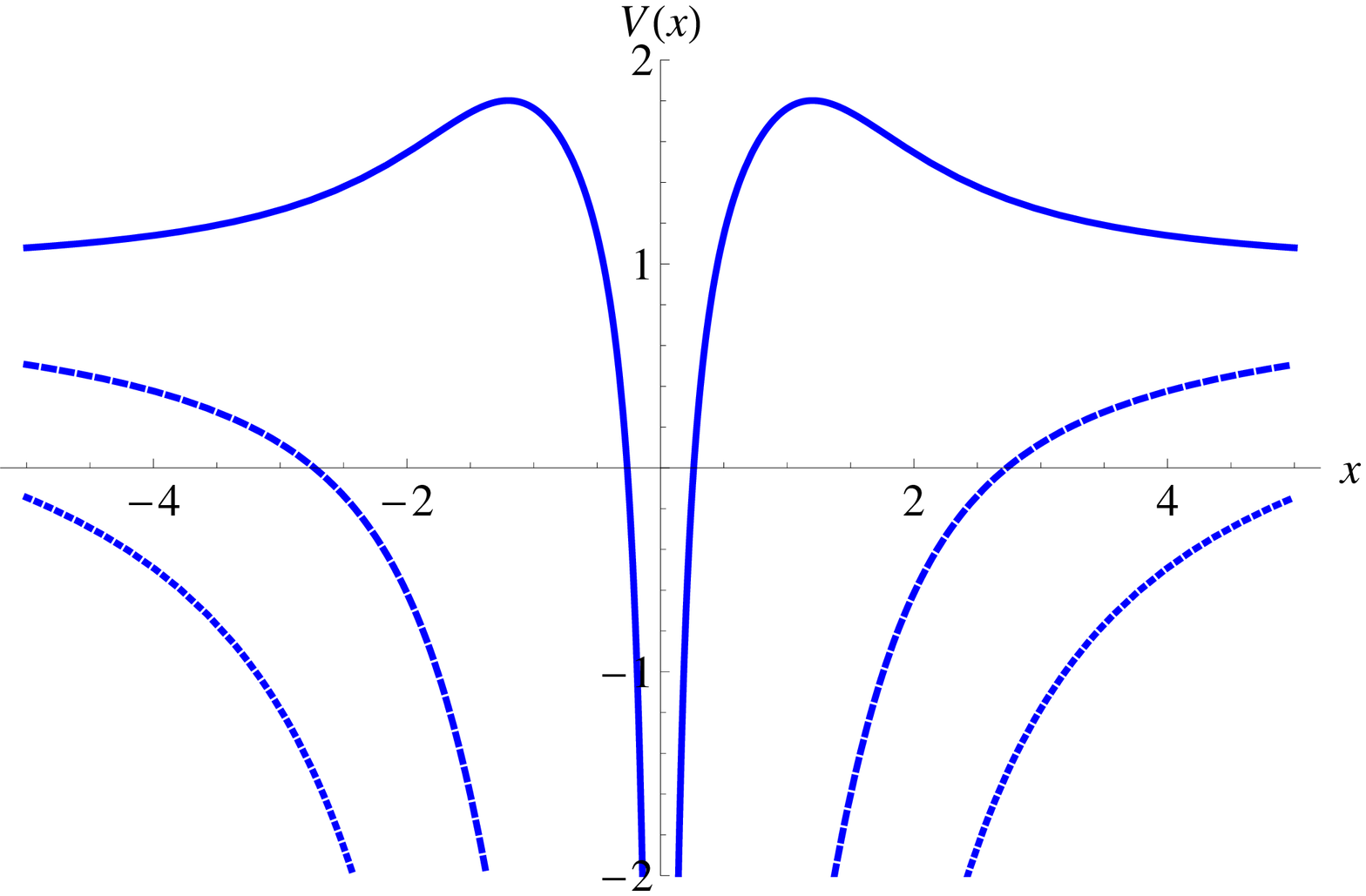}}
\caption{Potential for radial time-like geodesics. In each plot, the potential is shown for 3 different values of the number of charges $N_c$ (but the same $\delta_1$)}
\label{fig:TemporalRadial}
\end{figure}

The potential for radial time-like geodesics is shown in Fig. \ref{fig:TemporalRadial}. In this case, for large values of $r$, the geodesics feel an attractive potential that goes like $1-r_S/r$. The behavior near the wormhole depends on $\delta_1$ and is similar to the previous case. For $\delta_1<\delta_c$, the potential goes to $-\infty$ at the wormhole throat and the geodesics will cross to the other side. For $\delta_1=\delta_c$, the potential will be finite, and there will be stable orbits at the throat if $N_q<N_c$. If $\delta_1>\delta_c$, the potential will diverge to $+\infty$ and the geodesics will bounce back.

\section{Concluding Remarks}
We have shown that charged black holes for a $f(R,Q)$ gravity theory in the Palatini formalism possess a wormhole structure. This wormhole may have curvature divergences at its throat depending on the value of its charge-to-mass ratio. Using embeddings in Euclidean space, curvature divergences can be understood as a vertex at the wormhole throat. Null and time-like geodesics are shown to be able to cross the wormhole.

\section*{Acknowledgments}
Work supported by the Spanish grant FIS2011-29813-C02-02, the Consolider Program CPANPHY-1205388, the JAE-doc program of CSIC, the i-LINK0780 grant of CSIC, the NSFC (Chinese agency) grant No. 11305038, the Shanghai Municipal Education Commission grant for Innovative Programs No. 14ZZ001, the Thousand Young Talents Program, Fudan University, and by CNPq (Brazilian agency) project No. 301137/2014-5.

\end{document}